\documentclass[a4paper,twoside]{article}
\newcommand{\affil}[1]{$^{\rm #1}$}
\usepackage[authoryear]{natbib}
\bibpunct{(}{)}{;}{a}{}{,}
\usepackage{graphicx}
\date{} 
\newcommand{\kms}{\mbox{km\,s$^{-1}$}}
\title{\large\bf\flushleft 
A Damped Ly$\alpha$ Absorption-line System in an Apparent Void at Redshift 2.38
}
\author{\parbox{\textwidth}{\flushleft
\vspace{-0.5cm}
{\it Leith E. H. Godfrey\affil{A}, and Paul J. Francis\affil{A}\affil{B}}\\
\vspace{0.4cm}
{\small \affil{A}\,Research School of Astronomy and Astrophysics, Australian National
University, Canberra ACT 0200, Australia}\\
{\small \affil{B}\,Email: pfrancis@mso.anu.edu.au}}}
\begin{document}
\twocolumn[
\begin{changemargin}{.8cm}{.5cm}
\begin{minipage}{.9\textwidth}
\vspace{-1cm}
\maketitle
%
%
\small{\bf Abstract:} 

We study the contents of an apparent void in the distribution of Ly$\alpha$
emitting galaxies at redshift 2.38. We show that this void is not empty, but contains
a damped Ly$\alpha$ absorption-line system, seen in absorption against
background QSO 2138-4427. Imaging does not reveal any galaxy associated with
this absorption-line system, but it contains metals (Fe/H $ \sim -1.3$), and 
its large velocity range ($\sim 180$ \kms) implies a significant mass.


\medskip
\medskip
\end{minipage}
\end{changemargin}
]
\small

\section{Introduction}

It is now abundantly clear that high redshift galaxies are as strongly clustered
as those in the local universe \citep[eg.][]{gia98,ste00}. The topology of this 
clustering is not as clear, but there
is some evidence that high redshift galaxies, just like their local descendants,
lie in filamentary structures, separated by voids \citep[eg.][]{cam99,mol01,pal04,tur04}.

In the local universe, voids are surprisingly free of galaxies \citep[eg.][]{pee01}.
The few galaxies seen in voids are unusually blue, and tend to be centrally concentrated
\citep[eg.][and refs therein]{roi04}. Low column density Ly$\alpha$ forest QSO
absorption-line systems can also be found in voids \citep{shu96,pen04}. 
At higher redshifts, cold dark matter modelling predicts that galaxy biassing should be
much stronger than today \citep[eg.][]{fry96,bag98,teg98,bau99}. The {\em dark matter} 
underdensity in voids should thus be much less
than it is today. One might thus expect to find more or different objects in any voids.

\citet{pal04} mapped a $80 \times 60 \times 60$ co-moving Mpc region of the
high redshift universe, at $z = 2.38$ \citep[see also][]{fra04}. They showed that 
the distribution of Ly$\alpha$ emitting galaxies was not random. The void probability function was
significantly higher than that expected for randomly distributed galaxies on scales of 
5 --10 Mpc. The signal was dominated by a single large empty region just to the
north-west of their field centre.

These voids are deficient in Ly$\alpha$ emitting galaxies, but do they contain
anything else? Luckily, a luminous background QSO, LBQS 2138-4427, lies behind
the major void. \citet{fra93} showed that this QSO had a damped Ly$\alpha$
absorption-line system at redshift 2.383. This redshift places the damped
system clearly within the major Palunas et al. void, more
that 10 co-moving Mpc from any Ly$\alpha$ emitting galaxies. 

Is the QSO really passing through a void, or is the lack of Ly$\alpha$ emitting galaxies
close to it an artifact of small number statistics? In \citet{pal04} we generated mock
galaxy catalogues with the same total number of Ly$\alpha$ emitting galaxies as were observed. 
These mock
catalogues assumed a random distribution of galaxies, but allowed for differences in 
exposure time across our field (the QSO is in a region with above average exposure time). 
Less than
1\% of these random catalogues had no Ly$\alpha$-emitting galaxies within 10 projected
co-moving Mpc of the QSO sight-line. We therefore tentatively conclude that this sight-line is 
probing a region that is under-dense in Ly$\alpha$ emitting galaxies.

In this paper, we show that imaging fails to identify any galaxy candidate that could
be associated with this damped Ly$\alpha$ absorption-line system. We present
high resolution spectroscopy, which demonstrates, however, that the damped system
is substantially enriched with a wide variety of heavy elements. The velocity sub-structure
of the damped system also indicates the likely presence of significant mass. This void is
thus far from empty.

\section{Imaging}

The area around QSO 2138-4427 was imaged by \citet{pal04}, down to a Ly$\alpha$ flux limit of
$1.4 \times 10^{-16}{\rm erg\ cm}^{-2}{\rm s}^{-1}$, and down to (AB) broad-band
magnitude limits of $B=26.2$, $V = 25.3$, $R = 24.0$ and $I = 23.8$. Nothing
was seen within 10 arcsec of the QSO in any of these bands: it is an unusually clear
sight-line. Near-IR imaging was
obtained on the nights of September 6-7, 1998, using the CASPIR (Cryogenic
Array Spectrometer/Imager) camera on the Siding Spring 2.3m telescope.
A total of 4.5 hours exposure time was obtained in the H-band, reaching a (Vega) magnitude
limit of $H = 20.5$. Once again nothing was seen within 10 arcsec of the QSO.

\section{Spectroscopy}

\subsection{Observations and Reduction}

Our existing spectra of LBQS 2138$-$4427 \citep{fra93} were of too low 
a resolution and
narrow a wavelength coverage to seriously constrain the physical properties of
the gas. We therefore re-observed it using the University College London Echelle
Spectrograph (UCLES) on the Anglo-Australian Telescope (AAT). 
Observations were carried out on the nights of 2001 August 20 --- 23 and the 
total usable integration time was 43,200 sec. \citet{dod02} independently 
obtained high resolution spectra of this QSO, using the VLT Ultraviolet Visual 
Echelle Spectrograph (UVES). They kindly
allowed us to use their spectrum in this analysis. Both spectra have velocity
resolutions corresponding to a Doppler parameter $b = 6{\rm km\ s}^{-1}$.
Where they overlap, the two spectra agree very well.

The data were reduced using standard procedures, and set to the vacuum
heliocentric frame. 
The UVES spectrum has a higher signal-to-noise ratio, and a wider
wavelength coverage in the red: it was used for our analysis of most lines. The
C~IV doublet, however, was not covered by UVES so we used the UCLES
spectrum.

Absorption line properties were measured interactively using the XVOIGT
program \citep{mar95}. Upper and lower limits were derived interactively 
by varying parameters until residuals clearly exceeded the noise. In most
cases, possible line blending and/or saturation dominated noise as the major sourec of 
uncertainty.

\subsection{Results}

\begin{figure*}[h]
\begin{center}
\includegraphics[scale=1, angle=0]{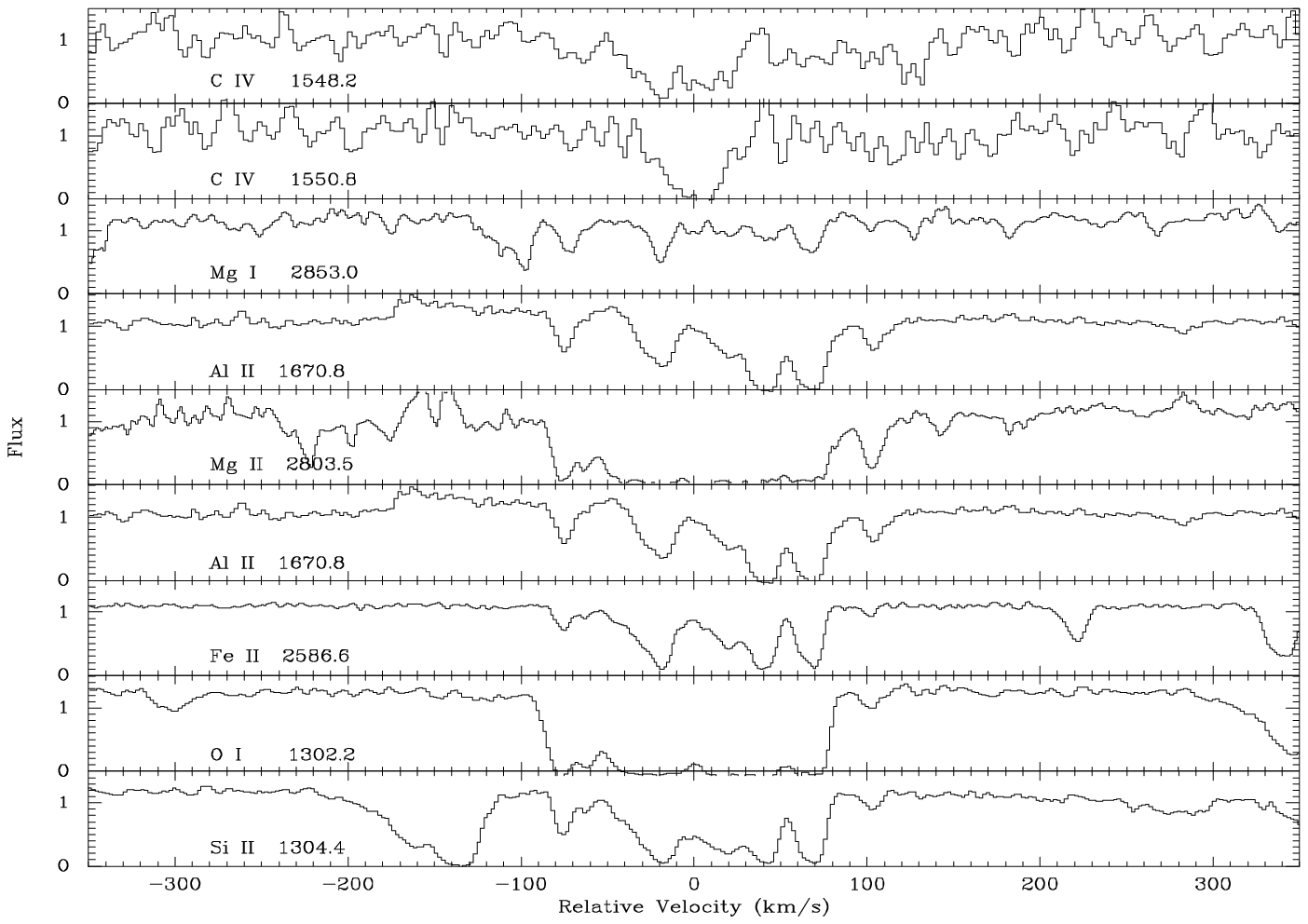}
\caption{
Spectra of the nine different transitions. The C~IV data are
from the UCLES spectrum, and the others are from the UVES
spectrum. Velocities are relative to a nominal 
redshift of 2.383.}\label{lineplot}
\end{center}
\end{figure*}

The damped Ly$\alpha$ system was clearly detected in a wide variety of
transitions (Fig~\ref{lineplot}). Low ionization transitions such as
O~I and Mg~II dominated, though a relatively weak detection
was made in C~IV.

\begin{table*}[h]
\begin{center}
\caption{Measured Parameters of Low-ionization Sub-components}\label{tab1}
\begin{tabular}{lccccccccc}
\hline Property & \multicolumn{9}{c}{Sub-component} \\
 ~  & 1 & 2 & 3 & 4 & 5 & 6 & 7 & 8 & 9 \\
\hline 
$v_{\rm rel}^a$ ($\kms$) 
       & -74   & -59 & -42 & -37 & -16 & 21 & 43 & 68 & 104 \\
$b^b$ ($\kms$)           
       & $< 6$ & $< 6$ & $< 7$ & $< 8$ & $< 10$ & $< 12$ & $< 10$ & $< 9$ & $< 6$ \\
 ~ & \multicolumn{9}{c}{Log(Column Density) (cm${-2}$)} \\

C~I    & $<12.6$ & $<12.7$ & $<12.7$ & $<12.7$ & $<12.7$ & $<12.7$ & $<12.7$ & $<12.7$ & $<12.7$  \\
C~II   & $14.4^{+0.9}_{-0.4}$ &  $14.2^{+0.6}_{-0.6}$ & \ldots & \ldots & \ldots & \ldots & \ldots & \ldots &  $13.8^{+1.2}_{-0.2}$ \\  
N~I    & $<13.3$ & $<13.7$ & $<13.5$ & $<13.7$ & $<14.5$ & $<14.1$ & $<14.5$ & $<14.0$ & $<13.6$ \\
O~I    & $15.0^{+1.9}_{-0.5}$ & $14.4^{+1.6}_{-0.2}$ & $14.2^{+1.7}_{-0.2}$ & $14.6^{+3.0}_{-0.4}$ & $14.9^{+1.0}_{-0.1}$ &
         $15.0^{+1.8}_{-0.2}$ & $15.2^{+1.8}_{-0.6}$ & $15.6^{+1.3}_{-0.6}$ & $13.1^{+0.3}_{-0.3}$  \\
Mg~I   & $11.6^{+0.3}_{\inf}$ & $11.1^{+0.3}_{\inf}$ & $< 11.2            $ & $<11.2$              & $12.0^{+0.2}_{-0.2}$ &
         $11.5^{+0.3}_{-0.2}$ & $11.6^{+0.3}_{-0.2}$ & $11.9^{+0.2}_{-0.2}$ & $11.4^{+0.4}_{-0.6}$  \\
Mg~II  & $13.4^{+0.8}_{-0.5}$ & $13.0^{+0.5}_{-0.3}$ & \ldots & \ldots & \ldots &
         \ldots & \ldots & \ldots & $13.1^{+0.6}_{-0.3}$  \\
Al~III & $12.1^{+0.4}_{-0.2}$ & $11.5^{+0.2}_{-0.5}$ & $11.0^{+0.5}_{-1.0}$ & $11.9^{+0.5}_{-0.3}$ & $12.4^{+0.2}_{-0.1}$ &
         $12.4^{+0.3}_{-0.2}$ & $13.4^{+1.6}_{-0.5}$ & $13.3^{+1.6}_{-0.5}$ & $12.0^{+1.1}_{-0.2}$  \\
Si~I   & $<12.0$ & $<12.0$ & $<12.0$ & $<12.0$ & $<12.0$ & $<12.0$ & $<12.0$ & $<12.0$ & $<12.0$  \\
Si~II  & $13.9^{+1.6}_{-0.6}$ & $12.8^{+0.8}_{-0.2}$ & $13.0^{+0.2}_{-0.3}$ & $13.0^{+0.4}_{-0.3}$ & $14.0^{+0.2}_{-0.1}$ &
         $14.0^{+0.2}_{-0.1}$ & $14.4^{+0.8}_{-0.6}$ & $13.9^{+0.3}_{-0.4}$ & $13.1^{+0.2}_{-0.1}$  \\
Si~III & $<12.6$ & $<12.3$ & $<12.7$ & $<12.5$ & $<15.4$ & $<14.8$ & $<13.0$ & $<16.3$ & $<13.6$  \\
Si~IV  & $<12.4$ & $<12.4$ & $<12.2$ & $<12.3$ & $<12.4$ & $<12.4$ & $<12.2$ & $<12.2$ & $<12.5$ \\
S~I    & $<12.7$ & $<12.7$ & $<12.7$ & $<12.7$ & $<12.7$ & $<12.7$ & $<12.7$ & $<12.7$ & $<12.7$ \\
Cr~II  & $<12.0$ & $<12.0$ & $<12.0$ & $<12.1$ & $<12.6$ & $<12.1$ & $<12.5$ & $<12.4$ & $<12.2$ \\
Fe~II  & $13.2^{+0.1}_{-0.2}$ & $12.9^{+0.1}_{-0.3}$ & $12.9^{+0.4}_{-0.2}$ & $13.2^{+0.4}_{-0.2}$ & $14.2^{+1.6}_{-0.4}$ &
         $13.7^{+0.3}_{-0.2}$ & $14.0^{+2.3}_{-0.1}$ & $14.0^{+0.6}_{-0.1}$ & $12.4^{+0.3}_{-0.1}$  \\
Fe~III & $<14.0$ & $<13.5$ & $<13.9$ & $<14.3$ & $<14.1$ & $<14.1$ & $<14.1$ & $<13.8$ & $<13.0$  \\
Ni~II  & $<12.6$ & $<12.6$ & $<12.6$ & $<12.6$ & $<13.0$ & $<12.4$ & $<12.8$ & $<12.6$ & $<12.6$  \\
Zn~II  & $<11.3$ & $<11.1$ & $<11.2$ & $<11.2$ & $<11.6$ & $<11.5$ & $<11.5$ & $<11.4$ & $<11.4$  \\
\hline
\end{tabular}
\medskip\\
$^a$Velocity of component relative to z = 2.383.\\
$^b$Velocity dispersion of individual sub-component.
\end{center}
\end{table*}

In the majority of transitions, the line breaks up into $\sim 9$
sub-components, with relative velocities spanning $178 {\rm km\ s}^{-1}$.
Each individual component is spectrally unresolved, or at best marginally
resolved. The measured properties of the components are shown in
Table~\ref{tab1}.
The neutral hydrogen content of each individual component cannot
be constrained, though the total neutral hydrogen content of the system is
log(N(HI)) = 20.5$\pm$0.1. In some cases (CII, MgII), the inner components blended into a single
absorption feature, and consequently an estimate of column
density for each inner component was not possible. However, an upper limit on
the combined column density could be determined in these cases. Several
ionic transitions did not produce observable absorption features (either due to
blending with unassociated lines, or low column density). For these
ions, only upper limits on column density were possible (assuming that their
velocity structure was the same as that of the other transitions). 
C~IV clearly has a different profile from the other (lower ionisation)
transitions and was best fit with two 
components at $v_{rel}$ = -2 and 14 kms$^{-1}$, each having 
N(CIV) $\sim 10^{14.2} {\rm cm}^{-2}$ with velocity
dispersion $ b \sim$ 10 -- 20 kms$^{-1}$. 

\subsection{Composition and Physical Properties of the Gas}

This system appears to be a very typical damped Ly$\alpha$
system. The predominance of low ionization species, and the narrowness
of the individual subcomponents suggests that they are cool and
relatively dense. We modelled a typical low ionization sub-component,
using Gary Ferland's CLOUDY code \citep{fer96}, and assuming photoionization
of a uniform density, plane parallel slab by a typical diffuse UV 
background \citep{haa96, sco00, sim04}. The line
ratios are consistent with the hydrogen being predominantly neutral
\citep{des04}, though we are unable to completely rule out partial
hydrogen ionization \citep[eg.][]{pro02b,vla01,izo01}. The modelling 
suggests densities of $\sim 10^{-3}$ -- $10^{-1} {\rm cm}^{-3}$ and 
a temperature of $\sim 10^4 K$. If we assume that the gas in these
components is mostly neutral, we infer mean metallicities [Fe/H], 
[Si/H], [O/H] and [Al/H] all of $\sim
-1.3$.  This is compared with the mean metallicity of $<[Zn/H]> \approx -1.15
\pm 0.15$ dex for DLAs at z $>$ 1.5 \citep{pro99}.

\section{Conclusion}

We conclude that this apparent void is not empty: it contains a very typical damped
Ly$\alpha$ absorption-line system. The non-detection of any galaxy counterpart
in our imaging is very consistent with this: only a tiny fraction of damped systems
have identifiable counterparts \citep[eg.][and refs therein]{bou04}. The
damped systems presumably either lie under the point spread function of the background QSO
or are below the detection limits.

The absorption-line spectra clearly show, however, that the gas in the damped
system is condensed into cool cloudlets, and that it is enriched in a wide
variety of elements. The large velocity dispersion between the components
further suggests that quite a large mass may be present.

We are not implying that all damped Ly$\alpha$ systems arise in voids. The inter-relationship
of damped systems and galaxies is clearly complex. On large scales, Lyman-break
galaxies and damped
systems seem on average to trace each other \citep{bou04}. In our field, there are two additional 
damped Ly$\alpha$ absorption-line systems, seen in the spectra of two different background 
QSOs. One of these lies within an apparent proto-cluster \citep{fra04b}, while the other
lies on the fringe of the major void \citep{fra01}. To add to the complexity, we are
delineating the void using Ly$\alpha$ emitting galaxies. Due to its high optical depth and 
easy destruction by dust, Ly$\alpha$ emission is not expected to correlate straightforwardly 
with any easily modellable property of a galaxy, such as its mass or star formation rate:
their distribution os thus hard to model. In addition, the 
apparent void is only 99\% significant. 

We tentatively conclude, however, that at least one damped system lies within an
apparent void, and that this damped system contains significant metal enrichment, and is
probably associated with considerable mass.

\section*{Acknowledgements}

We wish to thank Cedric Ledoux for making the UVES spectrum available to us.


\begin{thebibliography}{}

\bibitem[Bagla(1998)]{bag98} Bagla, J.~S. 1998, MNRAS, 299, 417

\bibitem[Baugh et al.(1999)]{bau99} Baugh, C.~M., Benson, A.~J., Cole, S., Frenk, C.~S.
\& Lacey, C.~G. 1999, MNRAS, 305, L21

\bibitem[Bouch\'{e} \& Lowenthal(2004)]{bou04} Bouch\'{e}, N. \& Lowenthal, J.D.
2004, ApJ, 609, 513

\bibitem[Campos et al.(1999)]{cam99} Campos, A., Yahil, A., Windhorst, R.~A., 
Richards, E.~A., Pascarelle, S., Impey, C., \& Petry, C. 1999, ApJL, 511, L4

\bibitem[D'Odorico et al.(2002)]{dod02} D'Odorico, V.,
Petitjean, P. and Cristiani, S. 2002, A \& A, 390, 13

\bibitem[Dessauges-Zavadsky et al.(2004)]{des04} 
Dessauges-Zavadsky, M., Calura, F. \& Prochaska, J.~X., 
	D'Odorico, S. and Matteucci, F. 2004, A \& A, 416, 79

\bibitem[Ferland(1996)]{fer96} Ferland, G.~J. 1996, ``Hazy, a Brief
Introduction to Cloudy'', University of Kentucky Department of Physics and
Astronomy Internal Report.

\bibitem[Francis \& Hewett(1993)]{fra93} Francis, P.~J. and Hewett,
P.~C. 1993, AJ, 106, 2587

\bibitem[Francis et al.(2004)]{fra04} Francis, P.~J., Palunas, P.,
Teplitz, H.~I., Williger, G.~M. \& Woodgate, B.~E. 2004, ApJ, 614, 75

\bibitem[Francis \& WIlliger(2004)]{fra04b} Francis, P.~J. \& WIlliger, G.~M. 2004,
ApJ 602, 77

\bibitem[Francis, Wilson \& Woodgate(2001)]{fra01} Francis, P.~J., Wilson,
G.~M. and Woodgate, B.~E. 2001, PASA, 18, 64

\bibitem[Fry(1996)]{fry96} Fry, J.~N. 1996, ApJ, 461, 65

\bibitem[Giavalisco et al.(1998)]{gia98} Giavalisco, M., Steidel, C.~C.,
Adelberger, K.~L., DIckinson, M.~E., Pettini, M. \& Kellogg, M. 1998, 
ApJ, 503, 543

\bibitem[Haardt \& Madau(1996)]{haa96} Haardt, F. \& Madau, P. 1996,
ApJ, 461, 20

\bibitem[Izotov et al.(2001)]{izo01} Izotov, Y.~I., Schaerer,
D. \& Charbonnel, C. 2001, ApJ, 549, 878

\bibitem[Mar \& Bailey(1995)]{mar95} Mar, D.~P. and Bailey, J.~G. 1995, PASA, 12,
239

\bibitem[M{\o}ller \& Fynbo(2001)]{mol01} M{\o}ller, P., \& Fynbo, J.~U. 
2001, A \& A, 372, L57

\bibitem[Palunas et al.(2004)]{pal04} Palunas, P., Teplitz, H.~I., Francis, P.~J., 
	Williger, G.~M. \& Woodgate, B.~E. 2004, ApJ, 602, 545

\bibitem[Peebles(2001)]{pee01} Peebles, P.~J.~E. 2001, ApJ, 557, 495

\bibitem[Penton, Stocke \& Shull(2004)]{pen04} Penton, S.~V., Stocke, J.~T. \& 
Shull, J.~M. 2004, ApJS, 152, 29

\bibitem[Prochaska \& Wolfe(1999)]{pro99} Prochaska, J.~X. \& Wolfe,
A.~M. 1999, ApJ, 121, 369
    
\bibitem[Prochaska et al.(2002)]{pro02b} Prochaska, J.~X., Howk,
J.~C., O'Meara, J.~M., Tytler, D., Wolfe, A.~M., Kirkman, D., Lubin,
D. \& Suzuki, N. 2002, ApJ, 571, 693

\bibitem[Roias et al.(2004)]{roi04} Roias, R.~R., Vogeley, M.~S.,
Hoyle, F. \& Brinkmann, J. 2004, ApJ, 617, 50

\bibitem[Scott et al.(2000)]{sco00} Scott, J., Bechtold, J.,
Dobrzycki, A. \& Kulkarni, V.P. 2000, APJS, 130, 67

\bibitem[Shull, Stocke \& Penton(1996)]{shu96} Shull, M.~J., Stocke, J.~T. \& 
Penton, S. 1996, AJ, 111, 72

\bibitem[Simcoe et al.(2004)]{sim04} Simcoe, R.~A. and Sargent,
W.~L.~W. and Rauch, M. 2004, ApJ, 606, 92
 
\bibitem[Steidel et al.(2000)]{ste00} Steidel, C.~C., Adelberger, K.~L.,
Shapley, A.~E., Pettini, M., Dickinson, M, \& Giavalisco, M. 2000, ApJ, 532,
170
 
\bibitem[Tegmark \& Peebles(1998)]{teg98} Tegmark, M. \& Peebles, P.~J.~E. 1998,
ApJL, 500, L79
 
\bibitem[Turnshek et al.(2004)]{tur04} Turnshek, D.~A., Rao, S.~M., Nestor, D.~B., 
Vanden Berk, D., Belfort-Mihalyi, M. \& Monier, E.~M. 2004, ApJL, 609, L53
 
\bibitem[Vladilo et al.(2001)]{vla01} Vladilo, G., Centuri{\' o}n, M., 
Bonifacio, P. \& Howk, J.~C. 2001, ApJ, 557, 1007

\end{thebibliography}
\end{document}